\documentstyle[11pt]{article}
\def\hybrid{\topmargin 0pt      \oddsidemargin 0pt
	\headheight 0pt \headsep 0pt
	\textheight 9in         
	\textwidth 6.25in       
	\marginparwidth .875in
	\parskip 5pt plus 1pt   \jot = 1.5ex}

\catcode`\@=11
\def\marginnote#1{}
\newcount\hour
\newcount\minute
\newtoks\amorpm
\hour=\time\divide\hour by60
\minute=\time{\multiply\hour by60 \global\advance\minute by-\hour}
\edef\standardtime{{\ifnum\hour<12 \global\amorpm={am}%
	\else\global\amorpm={pm}\advance\hour by-12 \fi
	\ifnum\hour=0 \hour=12 \fi
	\number\hour:\ifnum\minute<10 0\fi\number\minute\the\amorpm}}
\edef\militarytime{\number\hour:\ifnum\minute<10 0\fi\number\minute}

\def\draftlabel#1{{\@bsphack\if@filesw {\let\thepage\relax
   \xdef\@gtempa{\write\@auxout{\string
      \newlabel{#1}{{\@currentlabel}{\thepage}}}}}\@gtempa
   \if@nobreak \ifvmode\nobreak\fi\fi\fi\@esphack}
	\gdef\@eqnlabel{#1}}
\def\@eqnlabel{}
\def\@vacuum{}
\def\draftmarginnote#1{\marginpar{\raggedright\scriptsize\tt#1}}

\def\draft{\oddsidemargin -.5truein
	\def\@oddfoot{\sl preliminary draft \hfil
	\rm\thepage\hfil\sl\today\quad\militarytime}
	\let\@evenfoot\@oddfoot \overfullrule 3pt
	\let\label=\draftlabel
	\let\marginnote=\draftmarginnote
   \def\@eqnnum{(\theequation)\rlap{\kern\marginparsep\tt\@eqnlabel}%
\global\let\@eqnlabel\@vacuum}  }

\def\numberbysection{\@addtoreset{equation}{section}
	\def\theequation{\thesection.\arabic{equation}}}

\def\underline#1{\relax\ifmmode\@@underline#1\else
	$\@@underline{\hbox{#1}}$\relax\fi}
\def\titlepage{\@restonecolfalse\if@twocolumn\@restonecoltrue\onecolumn

\else \newpage \fi \thispagestyle{empty}\c@page\z@
	\def\thefootnote{\fnsymbol{footnote}} }

\def\endtitlepage{\if@restonecol\twocolumn \else  \fi
	\def\thefootnote{\arabic{footnote}}
	\setcounter{footnote}{0}}  
\catcode`@=12
\relax

\def\beq{\begin{equation}}
\def\eeq{\end{equation}}
\def\bea{\begin{eqnarray}}
\def\eea{\end{eqnarray}}
\def\bar{\overline}
\def\z{{\bar {z}}}
\def\nn{\nonumber}

\def\p{{\cal P}}

\def\N{ {\psi}}
\def\P{F}
\def\Pb{{\bar{F}}}

\def\m{\mu}

\def\rh{\rho}

\def\m{\mu}

\def\a{\alpha}
\def\b{\beta}
\def\g{\gamma}

\def\l{\lambda}
\def\demi{{1\over 2}}

\def\pz{\pa_z}
\def\pzb{\pa_\z }
\def\b{\beta}

\def\ee{\eea}
\def\be{\bea}

\def\p{\partial_{\bar z}}
 \def\pz{\partial_{\bar z}}
\def\pzb{\partial_{  z}}

\relax
\numberbysection
\hybrid
\begin{document}
\begin{titlepage}
\begin{center}
 \hfill	  DAMTP/96-50, PAR--LPTHE 96--15, YITP-96-47\\
[.5in] {\large\bf  SUPERSTRINGS  FROM THEORIES\\
WITH $N>1$ WORLD--SHEET SUPERSYMMETRY}\\
[.5in]
 {\bf Laurent Baulieu}\footnote{email address:
baulieu@lpthe.jussieu.fr} \\
{\it LPTHE, Universit\'es Paris VI - Paris VII, Paris, France}\footnote{
Universit\'es Paris VI - Paris VII, URA 280 CNRS,
4 place Jussieu, F-75252 Paris CEDEX 05, FRANCE.}\\
and \\
{\it Yukawa Institute for Theoretical Physics,
 Kyoto University, Kyoto 606, Japan}\\
{\bf   Michael B.  Green }\footnote{
M.B.Green@damtp.cam.ac.uk}\\
   {\it DAMTP,  Silver Street, Cambridge CB3
9EW, UK. }\\
{\bf Eliezer Rabinovici	 }\footnote{:eliezer@vms.huji.ac.il} \\
 {\it Racah Institute of Physics Hebrew University, Jerusalem,
Israel.}

\end{center}
\begin{quotation}
\noindent{\bf Abstract }
\noindent String theories with $(N,N')$ local world-sheet
supersymmetries are related to   each other by marginal
deformations.  This connects $N=1$ and $N=0$ theories in
which the  target-spaces are interpreted as   space-times,
$N=2$ theories in which the target spaces can be interpreted
as  world-volumes,   and theories with $N\ge 3$, in which the
 central charge vanishes -- theories with zero  target-space dimensions.

\end{quotation}
 \end{titlepage}

\newpage

%

\def\nn{\nonumber}

\newpage\null

 \section{ Introduction}
The recent progress in understanding non-perturbative features of
superstrings
 has revealed many striking relationships between  apparently
different
theories.   Theories with very different fundamental excitations
are
simply different expansions of the same  theory -- the solitonic
$p$-branes of
one theory, wrapped in a variety of ways around a compact space,  may
become
the fundamental strings of another.   Such dualities can relate
theories that
superficially have strikingly different properties --  for example,  
 a theory
in a particular
space-time
dimension can be related  to a theory in another dimension.
  An intriguing aspect of these dualities is that the r\^oles of the
world-volume
of a  $p$-brane and the target space are often not distinct -- the  
world-volume
of one theory may be the target space of another.      One aim of  
this paper
is to indicate how the distinction between the embedding space  and the
world-volume could  emerge  from a more fundamental formulation of  
the theory
by building upon an idea in \cite{nous}

It seems possible that  string theories with local $N=2$
supersymmetry \cite{ademollo1,ademollo2} have a special r\^ole to  
play in such
a reformulation of
string
theory.    Evidence for this arises from several sources.  (a)  The
$(2,0)$
heterotic version of such theories can have a target space with
$[1,1]$
signature which  may be interpreted as the {\it world-sheet} of the  
bosonic
string
\cite{green1}.  [ In order to avoid notational confusion the
left-moving and
right-moving supersymmetries will be denoted by curved brackets
$(.,.)$ while
the signature will be denoted by square brackets $[.,.]$ in the
following.]   As shown in  \cite{ooguri}  the target space may also
have
signature $[2,1]$ (the ambiguity in the target-space dimension is
associated
with an ambiguity in the choice of a null projector), in which case
it may be
interpreted as a bosonic version of the membrane world-volume
\cite{kutasov1,martinec1}.    (b)  The quantization of the  
heterotic   $(2,1)$
theories
was shown in \cite{ooguri} to  also
give theories with either $[1,1]$ or $[2,1]$ target spaces.  Those  
theories
with
$[1,1]$ target spaces  have precisely the field content needed
to describe
the world-sheets of  the critical superstring theories   \cite{kutasov1}
(see also \cite{ketov1}).  The
different  world-sheet theories  -- I,  IIA,  IIB, heterotic, as  
well as the
bosonic --
correspond simply to different choices of spin structures and
orbifolding in
various ways.   The case of  the $(2,1)$ theory with $[2,1]$
target-space
signature  has the field content  needed to describe the world-volume
of the
membrane of eleven-dimensional $M$-\lq theory'.   Thus, by making
changes in the boundary conditions   of the $(2,1)$ world-sheet
theory the
target space can be transformed into the world-volumes of any of  
the theories
(the
superstrings and $M$-\lq theory') that are related by duality
symmetries.   (c)
 The $N=2$ theory in its pure $(2,2)$ form describes self-dual
gravity  in
$[2,2]$ (or $[4,0]$) target-space dimensions while the heterotic
theories can
be viewed as projected  versions of self-dual gravity coupled to
self-dual
matter (field theories of this type are described in \cite{moore} and
references therein).  A point of
view
expressed in \cite{ooguri} emphasizes that the $(2,2)$ super-world-sheet
really has
four bosonic coordinates  --  a complex parameter $z$ conjugate to the
world-sheet hamiltonian and a complex parameter  $u$ conjugate to the
 $U(1)$ charge of the $N=2$
superconformal
group.   It can be interpreted as a brane of signature $[2,2]$,
which is the
same signature as the target space -- in fact, there is a symmetry
that
interchanges the  target space and the world-volume (see also  
\cite{rabinov3}).

The idea that the  world-sheet of one theory may be the target space
of another
is attractive although it seems fraught with  calculational and  
conceptual
problems.
For example, the calculation of a conventional string tree diagram would
require   second-quantized $N=2$ string field theory in a target  
space-time
that has the
appropriate topology.  In any
case, this view-point  still places emphasis on the distinction  
between the
world-volume  and the embedding space.
The observations in this paper may be taken
as an indication that both world-sheet and target space might be
described as different manifestations of the same underlying theory.
We will describe marginal deformations that  transform superstring   
theories
with  different target-space dimensions into each other.   Our
description will identify marginal operators that are used to deform the
target-space moduli.  From the point of view of the $(2,1)$
theory a subset of these theories are connected by deformations of the
world-sheet moduli.  A system in which a mapping between world-sheet and
target-space moduli can be identified is given in  \cite{rabinov1}.

 These deformations connect theories with target spaces that describe
space-time (such as the $(1,1)$ and $(1,0)$ superstrings) to  
theories with
target spaces that describe world-volumes (such as the $(2,1)$  
superstring) and
to theories which do not have any target space at all (theories  
with $N\ge 3$).

The initial observations (in section 2) will be based on considering
marginal
deformations of a $(2,2)$  theory.  Such deformations, which
correspond to
twisting various fields,  connect all critical string theories
with $(N,N')$ world-sheet supersymmetries (where $0 \le N, N' \le 2$).
In this
manner string theories with target spaces that may be interpreted as
world-volumes are
linked to conventional superstring theories by marginal deformations.
The initial
$(2,2)$ theory considered in section 2 consists of the usual set of
critical fields describing the $[2,2]$ target
space supplemented by a sector consisting of  a  \lq topological
package' of
matter and ghosts that  make cancelling contributions to the
central charge.
Such a package is defined to consist of four \lq BRST quartets', each
comprising  fermionic  fields with  conformal spins $(1,0)$ that will be
denoted  $(\lambda_1^\alpha,\rho_1^\alpha)$  ($\alpha = 1,2,3,4$)
and  their compensating bosonic fields  (which are also $(1,0)$  
fields)  which
will be
denoted $(\bar F_1^\alpha,F_1^\alpha)$. The fermionic fields
$(\lambda^\alpha,\rho^\alpha)$ might be related to Green--Schwarz  
fermions in a
manifestly  supersymmetric formulation of the theory.     In much  
the same way
as for the topological packages considered in \cite{nous} these  
fields have the
 BRST-exact action,
\be
I_{top} &=&
i\int d^2 z \sum_{\a=1}^{4} s(\Pb_1^\a \pz\rh_1^\a)\nn\\
&=&
\int d^2 z \sum_{\a=1}^{4}  (
i\l_1^\a \pz\rh_1^\a-\Pb_1^\a \pz\P_1^\a),\label{primedact}
\label{newprime}\ee
where the BRST transformations,  $s\rh_1^\a  = i\P_1^\a$, $s\P_1^\a  
 = 0$,
$s\Pb_1^\a = \l^\a _1$
and  $s\l^\a _1=0$ have been used.
   Evidently any number of such topological packages can be
added
to any theory without introducing any anomalies.

The process of  twisting the gravitino
ghosts in order to reduce the supersymmetry will also be described in
section 2. This changes the weight of one of the $(3/2,-1/2)$
pairs to $(1/2,1/2)$ thereby converting it into \lq matter',   
resulting in a
theory with reduced supersymmetry.   The anomaly-free condition can be
maintained if compensating twists are made on the $(1,0)$ fermions in a
topological package, $(\lambda_1^\alpha,\rho_1^\alpha)$, that  
transforms them
into conformal weight $(1/2,1/2)$ fields.  A version of the   
$(2,1)$ heterotic
theory is
obtained  after twisting
the right-movers in this manner.   A further twist on the left-movers
results in $(1,1)$ theories -- the type II superstrings.
In order to
proceed to theories with lower supersymmetry it is necessary to
add a second topological package, $(F_2^\alpha,\bar
F_2^\alpha);(\lambda_2^\alpha,\rho_2^\alpha)$ in the starting $N=2$
theory.  Then
the number of fields in the action is sufficient to allow  final
deformations giving rise to the $(1,0)$ heterotic string
and the $(0,0)$ bosonic string.

The presence of the extra topological matter fields suggests that the
$(2,2)$ theory
could itself be considered to have descended from a theory with more
supersymmetry.
Specifically, it will be shown in section 3 that there are  marginal
deformations of a critical theory with  the \lq large' $N=4$ local
world-sheet supersymmetry \cite{ademollo2,ohta1}
that correspond to various \lq twistings'
that take the theory to all possible critical
$(N,N')$ theories where $1\le N,N'  \le 4$.
The critical dimension of this theory
is zero -- in other words the total central charge of the ghosts
for this large local symmetry is $c^{(4)}_{ghosts} =0$.
 This means that the simplest
possible $(4,4)$ theory is one with no matter at all -- matter can
only be
added in the form of topological packages  with
$c^{(4)}_{matt} =0$.  In the first instance we will
consider the
purely gravitational $N=4$ theory in which there are no such extra
packages.   The   $N=4$ superalgebra consists of the Virasoro generators
together with seven generators of
an affine $SU(2)\times SU(2) \times U(1)$ and eight fermionic  
generators.  The
$(4,4)$
world-sheet theory therefore has eight bosonic generators for both  
left-movers
and right-movers.
Just as the $(2,2)$ string super world-sheet can be thought of as a  
$[2,2]$
world-volume when
the Kac--Moody parameters are included as bosonic dimensions,   the  
 $(4,4)$
world-sheet  can
be interpreted as an $[8,8]$ dimensional world-volume. This is  
suggestive of an
underlying
connection  with octonions \cite{martinec1}.

 Again the supersymmetry can be reduced by using compensating twists
on a gravitino
ghost  and other ghost fields, resulting in the  $N=3$ theory
(although this theory can
also be obtained simply by integrating out an anomaly free set
of fields, reflecting
the fact that the $N=3$ theory also has $c^{(3)}_{ghosts} =0$).
After two steps for both the right-movers and the left-movers the
$(2,2)$
theory is obtained, including the first set of topological matter
fields considered in section 2.
This leads, as before, to the string theories with $N,N'  =  1,2$.
To recover the theories with at least one $N=0$ chiral sector
requires the $N=4$ theory to contain   multiplets of  matter
fields in addition to the gravitational sector -- these necessarily
arise as $c=0$ topological packages.

Another reason for including additional topological packages arises from
considering chiral deformations of the  $(4,4)$ theory  to give heterotic
$(4,N)$ ($N\le 2$) theories.  There  are strong constraints imposed by
requiring that the $N=4$ fermionic ghosts have both left-moving and
right-moving components.  These can only be satisfied   if extra  
right-moving
topological packages of  fields are added to the initial theory, as  
will be
discussed in section 3.3.

Section 4 presents speculations as to how the
incorporation of these
extra topological packages   may arise naturally in theories with $N\ge
5$ local world-sheet supersymmetry.   Such theories  have vanishing  
central
charge as in the case of $N=4$,
but
they have a bigger ghost spectrum, determined from the antisymmetric
representations of $O(N)$ \cite{baulieu1}. After suitable
deformations
the extra ghosts give rise to
sufficiently many \lq matter' fields to extend the arguments of the
earlier
sections to include string theories with $N=0$ sectors.

The variety of target spaces obtained by deforming the same initial  
underlying
string theory (with $N\ge 4$)   include those that can also  be  
interpreted as
world-sheets of the underlying string theory itself.  We do not  
attempt to
answer the question  of whether the $[2,1]$ target space of the heterotic
$(2,1)$ string -- the suggested effective $M$-brane -- can also be  
interpreted
as a deformation of the  underlying world-volume theory to a  
three-dimensional
theory.   Such a relation between  theories in two and three  
dimensions seems
possible by analogy with  the equivalence  between the three-dimensional
level-$k$ Chern--Simons theory  in  the infinite $k$ limit and the
two-dimensional
$\phi *F$ system based on the same gauge group \cite{witten1,rabinov2}.
Deformations of either of these equivalent formulations  relate  
theories with
different  dimensions.

 Other somewhat different arguments have previously been advanced for
considering the embeddings of superstrings with lower supersymmetry  
within
those with higher supersymmetry \cite{berko1,giveon,englert}.

\section{N=2 world-sheet supersymmetry}
The basic idea of the construction is inspired by \cite{nous} where
the type II and heterotic strings were obtained as
different decompositions  of the same action.   This action, which   
was shown
to be a twisted version of a topological sigma model with bosonic  
and fermionic
coordinates  will be generalized to  a theory with local $N=2$
supersymmetry.  However, since we are now expecting to relate  
theories with
different target-space dimensions it will be necessary to twist  
some of the
ghost fields so that they become extra matter fields.

Let us review the various conformal components of the \lq standard'
critical $(2,2)$ supersymmetric superstring.   The fields of  extended
$N=2$ supergravity can be
described by the holomorphic Beltrami differential $\mu$, two
gravitini $\alpha_1,\alpha_2$, and a $U(1)$ gauge field $A$, together
with their antiholomorphic counterparts.   The associated ghost
sector consists of the
following fields.  The conformal weight-$(2,-1)$ fermionic ghosts for
general
coordinate invariance $(b,c)$ contribute $c=(-26,-26)$ to the central
charge.\footnote{The left-moving and right-moving central charges
will be  denoted  $c=(c_L,c_R)$.}  Two bosonic ghosts  for the
$N=(2,2)$ supersymmetry,
$(\beta^1,\gamma^1)$ and $(\beta^2,\gamma^2)$, each of
conformal-weight $(3/2,-1/2)$,
contribute a total of $c=(22,22)$.   The fermionic conformal-weight
$(1,0)$ ghosts
for the local $U(1)$ gauge subalgebra of the $N=2$ superconformal
symmetry,
$(f,g)$, contribute $c=(-2,-2)$.  Altogether the ghost sector
contributes $c_{ghosts}^{(2)}
= (-6,-6)$ to the central charge.  The compensating matter sector
includes four
bosonic dimensions, $P^\mu = \partial X^\m$, contributing  
$c=(4,4)$, and four
fermionic
spin-1/2
fields, $\N^\m $, contributing $c_{matt}^{(2)}=(2,2)$.  The target  
space of
this
theory is
therefore interpreted as  four-dimensional.  However, the $N=2$
supersymmetry
restricts this to even signature, $[0,4]$ or $[2,2]$\footnote{The  
target-space
coordinates
are naturally grouped in complex pairs ,  $\partial X^i, \partial  
X^{\bar i}$,
$\psi^i,\psi^{\bar
i}$ with $i,\bar i =1,2$.}  For the
present purposes
we will only consider the $[2,2]$ case.

This standard critical model can be generalized by adding extra
topological
matter. The particular topological packages of fields that  we will
consider  consist of the fields $(\bar
F_1^\alpha,F_1^\alpha)$;$(\lambda_1^\alpha,\rho_1^\alpha)$ and $(\bar
F_2^\alpha,F_2^\alpha)$;$(\lambda_2^\alpha,\rho_2^\alpha)$. The
$F_i,\bar
F_i$ are bosonic while $\lambda_i,\rho_i$ are fermionic.  These
fields  all
have conformal weight-$(1,0)$ and each bosonic pair contributes
$c=(2,2)$ for each
value of $\alpha$ while each fermionic pair contributes $c=(-2,-2)$
(and
$\alpha =1,\cdots,4$).  Thus, the  total central charge from this
sector
vanishes.   The field content of the $(2,2)$ theory is summarized  
in  table 1.

\vskip 0.5cm
\hfil\begin{tabular}
{ |c| |c |c |c|   |c|c |c| |c|c|c| } \hline
{}   & \multicolumn{3} {c||} {$N=2$}& \multicolumn{3} {c||} {$N=1$} &
\multicolumn{3} {c|}  {$N=0$} \\ \hline
& field &  weight& $c$    & field & weight&
$c$ & field   & weight & $c$ \\ \hline
- & $(b,c)$ &  $(2,-1)$ & -26  &$(b,c)$ &  (2,-1) &
 -26 &$(b,c)$ &  (2,-1) & -26\\ \hline
+  & $(\beta^1,\gamma^1)$ & (${3\over 2},-{1\over 2}$) & 11
&$(\beta^1,\gamma^1)$ &  $({3\over 2},-{1\over 2})$ & 11 &
$(\beta^1,\gamma^1)^T$ &  $({1\over 2},{1\over 2})$ & -1 \\ \hline
+ & $(\beta^2,\gamma^2)$ & $({3\over 2},-{1\over 2})$ & 11  &
$(\beta^2,\gamma^2)^T$  & $({1\over 2},{1\over 2})$ &
-1&$(\beta^2,\gamma^2)^T$  &  $({1\over 2},{1\over 2})$ &-1 \\ \hline
- & $(f,g)$& $(1,0)$ & -2  & $(f,g)$ & (1,0) & -2 & $(f,g)$ & (1,0) &
-2\\ \hline
+ &$P^\mu$& $(1,0)$ & 4   & $P^\mu$&  (1,0) & 4 &$P^\mu$&  (1,0) & 4
\\ \hline
-  &$\psi^\mu$& $({1\over 2},{1\over 2})$ & 2  &$\psi^\mu$& $({1\over
2},{1\over 2})$ & 2 &$\psi^\mu$& $({1\over 2},{1\over 2})$ & 2 \\
\hline
+ & $(\bar F_1^\alpha,F_1^\alpha)$ & $(1,0)$ & 8    & $(\bar
F_1^\alpha,F_1^\alpha)$ & $(1,0)$ & 8 &  $(\bar
F_1^\alpha,F_1^\alpha)$ & $(1,0)$ & 8 \\ \hline
-   &$(\lambda_1^\alpha,\rho_1^\alpha)$ & $(1,0)$ & - 8  &
$(\lambda_1^\alpha,\rho_1^\alpha)^T$&  $({1\over 2},{1\over 2})$  & 4
& $(\lambda_1^\alpha,\rho_1^\alpha)^T$&  $({1\over 2},{1\over 2})$ &4
\\ \hline
+&$(\bar F_2^\alpha,F_2^\alpha)$& $(1,0)$ & 8   & $(\bar
F_2^\alpha,F_2^\alpha)$ & $(1,0)$ &8 & $(\bar
F_2^\alpha,F_2^\alpha)$& $(1,0)$ & 8 \\ \hline
- &$(\lambda_2^\alpha,\rho_2^\alpha)$& $(1,0)$ & - 8  &
$(\lambda_2^\alpha,\rho_2^\alpha)$  &   $(1,0)$ & -8 &
$(\lambda_2^\alpha,\rho_2^\alpha)^T$  & $({1\over 2},{1\over 2})$ & 4
\\ \hline
\end{tabular}\hfil\break\noindent

{{\bf TABLE 1}: The fields of the holomorphic sector of the $N=2$
theory are listed together with their conformal weights and the total
contribution to the central charge, $c$, of the Virasoro algebra (
$\mu =1,\cdots,4$ and $\alpha=1,\cdots,4$).  The deformations that
reduce the supersymmetry to  $N=1$ and $N=0$ are indicated in the
last six columns.  The first column indicates the fermionic grading
of each field which is   unchanged by the deformations.}

\vspace{1.0cm}

The gauge-fixed lagrangian for this $(2,2)$ model can be written as
 \be I_B=I^{N=2}  + I_{top} , \ee
where the second term  is the  BRST-exact expression  (\ref{newprime})
involving only
the fields in the topological
package, which simply
decouple from the space of $\rho$-independent observables.  The
first term in the action is given by,
\be
\label{gaugeter}
&&I^{N=2}= \int d^2z  [b\pz c+ \b_1\pz\g_1 + \b_2\pz\g_2 +
f\pz g  \nn\\
&&   +
\sum_{\m=1}^{4} \left( \demi \pz X^\m\pzb X^\m -i\N^{\m  }\pzb \N^{\m }
\right) + {\rm complex\ conjugate}].  \ee
This action can also be interpreted (generalizing the expression in
\cite{nous} to the $N=2$ case)  as a topological $\sigma$ model in the
gauge defined by the conditions on the gauge fields,
\begin{equation}\label{gaugefix}
\mu =0, \qquad \alpha_i =0,\qquad A_z =0,
\end{equation}
(the last of these conditions gives rise to the  BRST-exact term  
$\int d^2z
s(fA_{\bar z}) = \int f\pz g$  in (\ref{gaugeter}))  and the holomorphic
conditions on the \lq matter' fields,
\be\label{holocon}
\partial X^1 + i \partial X^4 =0 = \partial X^2 + i \partial X^3, \qquad
\partial \rho_i^\alpha =0,
\ee
together with corresponding conditions in the anti-holomorphic
sector (strictly speaking these conditions are imposed in the usual  
gaussian
manner).

The topological sector decouples and thus appears to be redundant.
However,
after applying  appropriate marginal deformations which result in
twisting
some, and eventually all, of these extra fields  they will transmute
into matter
fields of  strings with less local supersymmetry.     The theories
defined by arbitrary deformations may not  be physically
relevant   but  the physically interesting
theories with $(2,1)$, $(1,1)$, $(1,0)$ and $(0,0)$ local
supersymmetry  are obtained  by specific finite deformations.

\subsection{Marginal deformations}
We will be interested in deformations which twist
various fields with conformal weights $\lambda^i$  in a way that  
maintains  the
vanishing of the total
central charge.   If these deformations are non-chiral they can be  
expressed as
the sum of  linear dilaton terms  in the action.  The action for  a  
set of $N$
such  fields, $\Phi^i$, is given by
\be\label{multidil}
\sum_{i=1}^N I_{k_i} =  \sum_{i=1}^{N} \int d^2z{1\over  {2\pi}} \sqrt{g}
\left( \pz
\Phi^i\pzb \Phi^i -{i \over 2}k^iR^{2}   \Phi^i\right).
\ee
where $R^{(2)}$ is  the world sheet curvature and the parameters   
$k^i$ define
the  central charge  $c=2 \sum_{i=1}^N  
\epsilon(6\lambda^{i2}-6\lambda^i+1)=
\sum_{i=1}^{N} (1-3{k^i}^2)$ (where $\epsilon =+1$ for bosons and  
$-1$ for
fermions).  Thus,
the  total central charge is unaltered by the presence
of such terms  if  $\sum_i{k^i}^2 =0$ (i.e., if $k^i$ is a null  
vector) and
the   $R^{(2)}$-dependent terms do  not then modify the total central
charge of the free bosonic system.

 The main way in which this formula enters in the following is in the
deformations that simultaneously  transform a bosonic  
$(\beta,\gamma)$ system
with $k_\beta \equiv k^5 = 2$ and the fermionic fields in a topological
package, $(\lambda^\alpha,\rho^\alpha)$, which have   
$k^\alpha_\lambda \equiv
k^\alpha = -1$ ($\alpha=1,\cdots,4$).   After the deformation the  
fields end up
with $k^\alpha=k^5 =0$.  A one-parameter family of deformed theories with
vanishing central charge can therefore  be defined by choosing continuous
deformations $k^5= - 2 k^\alpha = 2(1-m)$ with
 $0\le m \le 1$.  The Minkowski signature five-vector,  
$(k^5,k^\alpha)$, is
manifestly null at every intermediate value of $m$.
However, only for the special discrete values at the endpoints  
$m=0$ and $m=1$
  are these  assured  to be consistent unitary  string theories.

More generally, for chiral  fields, the twisting procedure is  
equivalent to a
change of field variables in which the relevant fields are  
multiplied by powers
of  a gravitino ghost, $\gamma$, as described in \cite{nous},  
\cite{baulieu5}
in the case of the $N=1$ theory.   Thus, for the example in the previous
paragraph the field redefinitions may be of the form,
\be
&&\gamma' = \gamma^{1/(1-2m)}, \qquad \rho' = \gamma^{- n} \rho,\nn \\
&&\beta' = \gamma^{-2m/1-2m} (1-2m) \left[\beta + n \gamma^{-1} \lambda
\rho\right], \qquad \lambda'= \gamma^n \lambda,
\ee\label{twistdef}
which ensure that $\int(\lambda^\alpha \bar \partial \rho^\alpha +  
\beta \bar
\partial \gamma) = \int(\lambda^{\prime \alpha}\bar \partial
\rho^{\prime\alpha} + \beta' \bar \partial \gamma')$.  A priori the  
parameters
$m$ and $n$ are independent,  taking the values $m=0$, $n=0$ for  
the undeformed
theory and $m=1$, $n=1$ for the final end-point of the marginal  
deformation.
The BRST transformations of the  fields at intermediate values of  
$m$ and $n$
follow simply from those of the original fields.  It is easy to see  
that the
condition for having a one-parameter family of anomaly-free theories that
interpolates between the end-point theories requires the choice,
\be\label{params}
n = - {m \over 1-2m},
\ee
so that $\gamma' = \gamma^{1-2n}$.
For generic  values of $m$ these theories are  probably   
ill-defined although
there may be well-defined theories at rational values of this parameter.
The BRST charge  $Q$ can be defined by this twisting procedure at  
any point
along the path of deformations starting from the initial theory    
(although the
explicit expression is rather lengthy and is omitted here).

\subsection{Marginal deformations of the $N=2$ theory.}
This procedure is now applied to the $N=2$ theory shown in the  
left-hand column
of table 1.  The first step consists of    simultaneously twisting the
right-moving  $(\beta^2,\gamma^2)$  and  
$(\lambda_1^\alpha,\rho_1^\alpha)$
fields.    These fields are transformed into  $(\beta^2,\gamma^2)^T$  and
$(\lambda_1^\alpha,\rho_1^\alpha)^T$ of
conformal weight $(1/2,1/2)$.    The left-moving sector  remains  as the
untransformed
model with $N=2$ supersymmetry.  After the transformation the field
content in the  right-moving sector (shown in   table 1)  consists of
the familiar N=1
ghosts --  the  $(b,c)$ and  $(\beta^1,\gamma^1)$ fields --  as well
as the
fermionic $(f,g)$ and bosonic $ (\beta^2,\gamma^2)^T$ fields.
Although these extra ghosts are  not needed in the usual $(1,1)$
theory they  arise here because of the pairing of left-moving and
right-moving scalar fields in the  $(2,1)$ theory.  Thus, the ($\bar
f,\bar g)$  ghosts for the $U(1)$ symmetry are paired with the
right-moving  fields, $(f,g)$.    Right-moving  $N=1$ local
supersymmetry requires, in addition, the presence of the bosonic
right-moving fields  $(\beta^2,\gamma^2)^T$. The deformed theory thus
possesses the appropriate
$(2,1)$ supersymmetry ghost system (as in \cite{ooguri})

The matter fields of this $(2,1)$ model include those of the original
$(2,2)$ theory together with
$(\bar F_1^\alpha,F_1^\alpha)$ which may now be  interpreted as eight
real
bosonic fields adding $[0,8]$  right-moving target-space dimensions  
to the
$[2,2]$ already present.   Consistency requires these to be  
compactified on a
euclidean self-dual lattice, namely, the  $E_8$ lattice.  These  
bosons can also
 be
written in terms of 16  $(\demi,\demi)$ fermions which can be  
grouped with the
eight right-moving fermions, $(\lambda^\alpha_1,\rho_1^\alpha)$,    
to give more
general target-space theories by appropriate choices of spin  
structures (or GSO
projections) \cite{kutasov1}.
These are right-moving bosonic  fields  of the usual $(2,1)$   
heterotic case.

 The four
fermionic matter fields of the $N=2$ theory, $\N^\m $,
are similarly supplemented by $(\lambda_1^\alpha,\rho_1^\alpha)^T$,
giving a total of twelve fermions which are superpartners of the
bosonic matter fields. The target-space theory
is therefore described as a heterotic theory with $[2,2]$ signature
for the left-movers and $[10,2]$ for the right-movers.   As in
\cite{ooguri,kutasov1}  the
gauge constraint on the $U(1)$ current
leads to a null reduction that eliminates one of the time directions
and results in a target-space  that is either  $[1,1]$ or $[2,1]$.   
This null
reduction pairs the $(f,g)$ ghosts in a topological package with  
one of the
time-like $P^\mu$ fields and either a space-like $P^\mu$ or
The topological package  consisting of $(\bar
F_2^\alpha,F_2^\alpha)$;$(\lambda_2^\alpha,\rho_2^\alpha)$ descends
untouched  to the $N=(2,1)$ theory.

It is now obvious that a similar marginal deformation of the
left-moving sector will change the theory with  $(2,1)$ supersymmetry
to one with $(1,1)$ supersymmetry.  This defines the usual type II  
theories
in an enlarged $[10,2]$ target space together with   the ghosts   
$(f,g)$ and
$(\beta^2,\gamma^2)^T$.  These ghosts are associated with  gauge
constraints  that can be used  to reduce  to the usual critical  
$[9,1]$  theory
by forming two BRST topological quartets of fields -- the first  
consisting of
$(f,g)$ with two of the spin-$(1,0)$ bosonic fields and the second of
$(\beta^2,\gamma^2)^T$ with two of the spin-$(1/2,1/2)$ fermionic fields.
The $(\bar F_2^\alpha,F_2^\alpha)$;$(\lambda_2^\alpha,\rho_2^\alpha)$
topological package is again untouched and is simply  appended to
the  $N=(1,1)$ model.

Strictly speaking, if the twisting is carried out independently for the
left-movers and the right-movers the matter fields describe  
compactified target
spaces.  More generally, BRST-exact terms can be added to the  
action  that
relate the two world-sheet directions, leading to extra zero modes  
associated
with uncompactified dimensions.  For example, denoting the  
left-moving and
right-moving bosonic fields $(F_1,\bar F_1)$ by $(F_L,\bar F_L)$  
and $(F_R,\bar
F_R)$, gives a first-order action,
\be
\label{leftright}
\int d^2z \left(F_L \partial \bar F_L + F_R \bar \partial \bar F_R  
+ F_L F_R
\right),
 \ee
where the last term (which has ghost number 2) is of the form $\int  
d^2z s(F_L
\rho_R)$.
Integrating out $F_L$ and $F_R$ leaves the second-order action,  
$\int d^2z \bar
F_L\partial \bar \partial \bar F_R$.   If $F_L^\alpha$ and  
$F_R^\alpha$ are
real their zero modes describe extra  dimensions of signature  
$[4,4]$.   If
they are to describe purely spatial dimensions it is necessary to  
continue them
to complex values and impose the reality condition, $\bar F^*_R =  
\bar F_L$.

The presence of  extra topological sectors allows  further
twistings that reduce the $(1,1)$ theory to $(1,0)$.  This is
obtained by the  marginal deformation that twists
$(\beta^1,\gamma^1)$  into
$(\beta^1,\gamma^1)^T$ and simultaneously twists the fermionic
components of the topological package  from $(\lambda_2,\rho_2)$
into $(\lambda_2,\rho_2)^T$.   The right-moving ghosts now comprise
$(b,c)$
together with  two BRST quartets consisting of the bosonic
fields $(\beta^1,\gamma^1)^T$ and $(\beta^2,\gamma^2)^T$ together
with two of the fermionic fields of the same conformal weight,
$(\lambda_1^4,\rho_1^4)^T$, $(\lambda_2^4,\rho_2^4)^T$.
This leaves twelve  right-moving Weyl fields,
$(\lambda_1^i,\rho_1^i)^T$ and
$(\lambda_2^i,\rho_2^i)^T$ (where $i=1,2,3$).  In addition there  
are eight
right-moving chiral bosons, $(\bar F_2^\alpha,F_2^\alpha)$ as well as
the four
original  right-moving fermions, $\N^\m $, in the matter sector.
This  gives a total of eight right-moving  bosons and sixteen
fermions, which corresponds to  a version of the usual content of the
right-moving
sector of the heterotic string\footnote{This content is conventionally
expressed in terms of 16 chiral bosons or 32 chiral fermions.}.

Finally, the conformal  spin of the left-moving gravitino ghosts, $(\bar
\beta^1,
\bar\gamma^1)$, may be deformed together with the second  topological
package  of left-moving fields to give the $(0,0)$ bosonic string
theory.  After bosonising the sixteen left-moving and right-moving
fermions they can be identified with eight toroidally compactified
bosons. Together with the bosonic fields  $(\bar
F_1^\alpha,F_1^\alpha)$, $(\bar F_2^\alpha,F_2^\alpha)$  and their
antiholomorphic partners, these make up an extra sixteen dimensions
of the 26-dimensional target space of bosonic string theory.

\section{Deformations of the large $N=4$ theory.}
The presence of the extra topological packages  in the $(2,2)$ theory
suggests that it might be embedded naturally in a theory with higher
supersymmetry.  We will therefore   consider  the purely gravitational
theory with the \lq large' $N=4$ local superconformal symmetry
\cite{ademollo2}.

The generators of the large $N=4$ superconformal algebra \cite{ohta1}
consist of
the holomorphic and anti-holomorphic components of the
energy-momentum tensor, the four fermionic supercurrents, the  seven
generators of a $U(1) \times O(4)$ (or $U(1)\times SU(2)\times
SU(2)$) affine algebra and four more fermionic \lq internal symmetry'
generators.   Each of these generators  can be associated with a
two-dimensional gauge field.    The purely gravitational action for
this $N=4$ theory may then be deduced by setting these gauge fields
to zero by a gauge choice that is implemented in a BRST invariant
manner (again generalizing the discussion in the $N=1$ case).

The ghosts for the various gauge symmetries will be denoted by
$c $ (for reparameterizations), $\gamma^a$ (for the four local
supersymmetries with $a=1,2,3,4$),  $c^{+i}$, $c^{-i}$  (for the
$SU(2)\times SU(2)$ gauge symmetry, where $i=1,2,3$), $\epsilon^a$
(for
the four additional fermionic internal symmetries)  and
$g$ (for the local $U(1)$ symmetry).  Ghost fields  of integer weight
are
fermions while the rest are bosons.
Correspondingly, in the generalized conformal gauge (in which the
gauge fields vanish)  the antighosts are $b$, $\beta^a$, $ b^{\pm
i}$, $\delta^a$     and $f$  which are respectively associated  with
$c$, $\gamma^a$, $c^{\pm i}$,  $\epsilon^a$ and $g$.

The ghost action with the large $N=4$ worldsheet invariance is given
by
\be\label{mother}
I^{N=4}  &=&
\int  d^2z  \left(b\p c +\sum_{a=
1}^4\quad\left( \beta^a \partial_{\bar z} \gamma^a + \delta^a
\partial_{\bar z} \epsilon^a \right)\right.\nn\\
&&\left. +\sum_{i= 1}^3\quad\left(b^{+i}\p c^{+i}+
 b^{-i} \partial_{\bar z} c^{-i}\right)+ f \partial_{\bar z} g\right).
\ee
This can be expressed as the s-exact action
\be
I^{N=4} =
\int d^2z\sum_{\rm all sectors}  s\left( B_{({\cal A})} {\cal A}
\right),
\ee
where ${\cal A} $ stands for the $\z$ components of the gauge
fields with  corresponding antighosts
$B_{(\it {A})}$.
This action enforces the gauge conditions
${\it A} =0$ in a BRST invariant manner.

This ghost system is  anomaly free.   The  central charges of the
pairs $(b,c)$,
 $(\beta^a,\gamma^a)$,
$(\epsilon^a,\delta^a)$,
$(b^{+i},c^{+i})$,
$(b^{-i},c^{-i})$
and
$(f,g)$ give contributions to the central charge of the Virasoro
algebra equal to
$-26$,  $4\times 11=44$, $4\times (-1)=-4$,
$3\times (-2)=-6$, $3\times (-2)=-6$
and $-2$, respectively.  The  total central charge, $c^{(4)}_{ghost}$,
therefore vanishes
as commented on earlier.  This means that the only other fields that
can be added to this system must be topological packages.

The relationship of the large $N=4$ theory to the one with the \lq
small' $N=4$ symmetry \cite{ademollo2}
is very simply stated in terms of these fields.
 The fields $(b,c)$, $(\beta^a,\gamma^a)$ and $(b^{+i},c^{+i})$ are
the ghost fields of the small algebra and have a total central charge
of 12.
The remaining eight commuting ghosts and the eight anticommuting
ghosts  in (\ref{mother}) are then interpreted as a \lq matter'
multiplet of the small $N=4$ theory with central charge $-12$.  This
is the origin of the statement that the critical dimension of this
theory is $-8$.  In a certain sense it is a theory in which the
r\^oles of ghosts and matter have been interchanged.  [A very different
interpretation of the small $N=4$ theory is given in  \cite{siegel1}.]

One curiosity is that the large $N=4$ theory can also be interpreted
as a theory with
$N=3$ supersymmetry.  This can be seen by integrating out an
anomaly-free subset of fields consisting of one of the gravitino
ghost pairs $(\beta^4,\gamma^4)$,  one of the $SU(2)$
ghost packages, $(b^{-i},c^{-i})$, three of the internal symmetry
fermion ghost pairs, $(\epsilon^a,\delta^a$) with $a=1,2,3$, and the
$U(1)$ ghost pair, $(f,g)$.  These fields contribute respectively
$11$, $-6$, $-3$ and $-2$ to the central charge.  The remaining
fields consist of
$(b,c)$,  $(\beta^i,\gamma^i)$,
$(b^{+i},c^{+i})$, with $i=1,2,3$, and the remaining fermionic ghost
system, $(\epsilon^4,\delta^4 )$.   This is precisely the content of
the theory with
 $N=3$ local supersymmetry, which again has vanishing central charge
in the gravitational sector. Its invariances  include  a  local
$SU(2)\sim O(3)$ gauge symmetry and    a local fermionic symmetry of
rank one.
There is no way to reduce the supersymmetry any further simply by
integrating out fields.

However, as with the $(2,2)$ theory, it
can be reduced by twisting fields in a manner that preserves the
vanishing of the total central charge.   These deformations are
illustrated in table 2.

\noindent\begin{tabular}
{ |c||c|c|c|   |c|c|c| |c|c|c| |c|c|c|  } \hline
{} &\multicolumn{3}{c||} {$N=4$}   & \multicolumn{3} {c||} {$N=3$}&
\multicolumn{3} {c|} {$N=2$}  \\ \hline
 & field   & weight& $c$ & field   & weight&
$c$& field  & weight &c \\ \hline
-  &$(b,c)$ &  $(2,-1)$ & -26 & $(b,c)$ &$(2,-1)$ & -26  &$(b,c)$ &
(2,-1) &
 -26  \\ \hline
+  & $(\beta^1,\gamma^1)$ &(${3\over 2},-{1\over 2}$) & 11
&$(\beta^1,\gamma^1)$ &  $({3\over 2},-{1\over 2})$ & 11
&$(\beta^1,\gamma^1)$ &  (${3\over 2},-{1\over 2}$) & 11   \\ \hline
+ & $(\beta^2,\gamma^2)$ &$({3\over 2},-{1\over 2})$ & 11 &
$(\beta^2,\gamma^2)$ &  (${3\over 2},-{1\over 2}$) & 11  &
$(\beta^2,\gamma^2)$   & (${3\over 2},-{1\over 2}$) & 11 \\ \hline
+ &$(\beta^3,\gamma^3)$ & $({3\over 2},-{1\over 2})$ & 11 &
$(\beta^3,\gamma^3)$  & (${3\over 2},-{1\over 2}$) & 11 &
$(\beta^3,\gamma^3)^T$ & $({1\over 2},{1\over 2})$ & -1 \\ \hline
+ &$(\beta^4,\gamma^4)$ & $({3\over 2},-{1\over 2})$ & 11 &
$(\beta^4,\gamma^4)^T$  &  $({1\over 2},{1\over 2})$ & -1 &
$(\beta^4,\gamma^4)^T$  &  $({1\over 2},{1\over 2})$ & -1 \\ \hline
 - &$(b^{-i},c^{-i})$ & $(1,0)$ & -6 & $(b^{-i},c^{-i})$ & $(1,0)$ &
-6 &
$(b^{-i},c^{-i})^T$ & $({1\over 2},{1\over 2})$ & 3 \\ \hline
- &$(b^{+i},c^{+i})$ &  $(1,0)$ & -6 &
$(b^{+i},c^{+i})^T$   &  $({1\over 2},{1\over 2})$ & 3 &
$(b^{+i},c^{+i})^T$   & $({1\over 2},{1\over 2})$ & 3 \\ \hline
 + &$(\delta^1,\epsilon^1)$ & $({1\over 2},{1\over 2})$ & -1  &
$(\delta^1,\epsilon^1)$ & $({1\over 2},{1\over 2})$ & -1&
$(\delta^1,\epsilon^1)$ & $({1\over 2},{1\over 2})$ & -1\\ \hline
+ & $(\delta^2,\epsilon^2)$ & $({1\over 2},{1\over 2})$ & -1
&$(\delta^2,\epsilon^2)$ & $({1\over 2},{1\over 2})$ & -1
&$(\delta^2,\epsilon^2)$ & $({1\over 2},{1\over 2})$ & -1\\ \hline
+ &$(\delta^3,\epsilon^3)$ &  $({1\over 2},{1\over 2})$ & -1
&$(\delta^3,\epsilon^3)$ &  $({1\over 2},{1\over 2})$ & -1
&$(\delta^3,\epsilon^3)$ &  $({1\over 2},{1\over 2})$ & -1\\ \hline
+ &$(\delta^4,\epsilon^4)$ &  $({1\over 2},{1\over 2})$ & -1 &
$(\delta^4,\epsilon^4)$  & $({1\over 2},{1\over 2})$ & -1 &
$(\delta^4,\epsilon^4)^T$  &   $(1,0)$ & 2\\ \hline
 - &$(f,g)$& $(1,0)$ & -2  &
$(f,g)^T$& $({1\over 2},{1\over 2})$  & 1 &  $(f,g)^T$  &  $({1\over
2},{1\over 2})$  & 1     \\ \hline
\end{tabular}\break\noindent

{{\bf TABLE 2}: The fields of the holomorphic sector of the $N=4$
theory and the deformations that reduce the supersymmetry to $N=3$
and $N=2$.  The
purely gravitational $N=4$ model is taken as the starting point (with
$i=1,2,3$).  The field assignments in the $N=2$ theory differ from
those in table 1 due to ambiguities in the choice of deformations.}

\subsection{Deformation to $N=3$ plus $1$ topological package.}
Instead of integrating out the anomaly-free subset of fields in   
passing from
the $N=4$ to the $N=3$ theory they can be twisted into a set of  
cancelling
weight-$(1/2,1/2)$ bosonic and fermionic fields. Thus,  twisting one
pair weight-$(3/2,1/2)$  gravitino ghost pair, $(\beta^4,\gamma^4)$, to
weight-$(1/2,1/2)$ bosonic fields $(\beta^4,\gamma^4)^T$ changes the
central charge by $-12$.  This change may be compensated by twisting
three of the six weight-$(1,0)$ fermionic ghosts,$(b^{+i},c^{+i})$
associated with one of the internal $ SU(2)$ gauge symmetries into
three fermionic weight-$(1/2,1/2)$ fields,    $(b^{+i},c^{+i})^T$.
Each of these fields has its contribution to the central charge
increased from $c=-2$ to $c=1$. In addition,  the weight-$(1,0)$
fermionic ghost of the $U(1)$ symmetry, $(f,g)$, can be  transformed
into a fermionic weight-$(1/2/,1/2)$ field, $(f,g)^T$, also changing
its central charge by $+3$.
The overall increase in the central charge due to the twisting
of the $(1,0)$ ghosts is $12$, which cancels the
decrease from twisting $(\beta^4,\gamma^4)$.   These twistings
therefore retain the vanishing central charge, $c^{(3)}_{ghost} =0$,  and
reduce the
supersymmetry, which gives a path to the $N=3$ theory that is  
different from
that based on integrating out an anomaly-free multiplet of  fields.

The field content of this $N=3$ theory consists of the  anomaly-free
set of ghost fields --  $(b,c)$, $(\beta^i,\gamma^i)$,
$(b^{+i},c^{+i})$, $(\epsilon^4,\delta^4)$. In addition to this
standard content of the $N=3$ model the remaining fields can be
grouped together into topological packages consisting of a pair of
opposite statistics  $(1/2,1/2)$ fields such as
${(\epsilon^i,\delta^i);(b^{+i},c^{+i})^T}$ and
${(\beta^4,\gamma^4)^T;(f,g)^T}$.    Although there is a great deal
of ambiguity in the choice  of marginal deformations  the  above
choice is the one which makes the local  $SU(2)$ invariance manifest.

\subsection{Deforming to $N\le 2$.}
 This process can obviously be repeated to obtain a theory with $N=2$
supersymmetry as illustrated by the transformations in table 2.  The
field representation  of this $N=2$ theory differs from that used in
section 2 but a further set of twists would transform these into the
same definitions as in table 1.    [A continuous algebraic  
deformation of the
$N=2$ algebra to the large $N=4$ algebra appears in \cite{giveon}.]

This demonstrates how  a string theory with the large
$N=4$ symmetry (\ref{mother})  can be interpreted as   an $N=2$ critical
superstring theory,  together with a purely
topological sigma model based on four fermionic and four bosonic
degrees
of freedom.  String theories with $N=1$ supersymmetric sectors (the
$(2,1)$ and $(1,1)$ theories) can now be described by the same
deformations of $(\beta^2,\gamma^2)$ and the fermionic topological
fields that were discussed in section 2.
However, only the topological package  of fields  $(\bar
F^\alpha_1,F^\alpha_1);(\lambda^\alpha_1,\rho^\alpha_1)$  is   
obtained in the
$N=2$ theory
whereas we saw in section 2 that in order to deform the theory into  
one with
at least one $N=0$ sector a second set of similar fields is needed.

\subsection{Heterotic $N=4$ theories}

 Up to this point we have assumed for simplicity  that the  
deformations of the
$(4,4)$ theory are non-chiral so that there are no essentially new issues
relating to  the constraints  between left-movers and right-movers  
that arise
in heterotic theories.  In considering  deformations that take the  
$(4,4)$
theory to  $(4,2)$, $(4,1)$ and $(4,0)$ it is necessary to add  
extra fields in
order to ensure that the fermionic ghosts are non-chiral.  This is  
the analogue
of the requirement in the $(2,0)$  theory that the $(f,g)$ pair has  both
left-moving and right-moving components as shown in table 1.    In  
that case
the presence of the  extra $c=-2$ right-moving  fields led to  a total of
$c=-28$  from the ghosts  that had to be cancelled by the matter.  In the
$(2,1)$ theory the $(f,g)$ pair were accompanied by  $N=1$  
superpartners and
the matter sector had $c=12$.

The $N=4$ ghost system has a set of  fermionic  ghosts that  
contribute $c=-40$
(the $-$ fields in  table 2).  In deforming to the $(4,2)$ theory  
we now want
to leave the right-moving fermionic ghosts $(f,g)$ and $(b^{+i},c^{+i})$
undeformed ({\it unlike} the deformation illustrated in table 2).    
  In this
case it is necessary to append exactly two right-moving   
topological  packages
to the original pure $(4,4)$ theory.  Then the deformation that twists
$(\beta^3,\gamma^3)$ and $(\beta^4,\gamma^4)$ can be chosen so as to
simultaneously twist the fermionic fields in these two packages  
instead of
twisting  $(f,g)$ and $(b^{+i},c^{+i})$.   The resulting ghost  
fields of the
right-moving sector  now consist of  the original $N=4$ ghosts, but  
with  two
gravitino ghost pairs twisted,  giving  a total of $c=-24$.  It is  
easy to see
that these fields fall into representations of the $N=2$  
supersymmetry.  The
\lq matter' fields come from the two  deformed
packages which each contribute $c=12$.  These form four $N=2$  
supermultiplets.
The resulting theory may therefore be thought of as a $(4,2)$  
theory with a
target space that has no space-time dimensions but  with   
right-moving  fields
of the $N=2$ theory compactified on a self-dual lattice together  
with extra
space and time coordinates that combine with extra ghosts into   
topological
BRST quartets.

This procedure extends to the $(4,1)$ theory  if three right-moving  
topological
packages are added to the initial $(4,4)$ theory.   The  
deformations are chosen
to twist three of the $(\beta,\gamma)$ pairs and the fermionic  
fields in the
packages.  The ghosts in the right-moving sector now contribute a  
total of
$c=-36$ which is cancelled by the $c=36$ \lq matter'  from the twisted
topological packages.  This right-moving matter  can be arranged in  
a variety
of manners consistent with modular invariance.  In particular, this  
theory  has
a target space that can be interpreted as that of a zero-dimensional
world-volume embedded in a ten-dimensional target space (again  
there are  extra
space and time dimensions that combine with ghosts into topological BRST
quartets).  This is intriguingly suggestive of the  \lq  
world-volume' of the
$D$-instanton embedded in the target space of  type II superstring  
theory.

In the $(4,0)$ theory the right-moving fermionic ghosts contribute  
$c=-40$.
The obvious extension of the above procedure requires  a minimum of four
right-moving topological packages,
$(\bar F_r^\alpha,F_r^\alpha);(\lambda_r^\alpha,\rho_r^\alpha)^T$
($r=1,\cdots,4$),  to be added to the pure $N=4$ theory.   Twisting  
the four
gravitino ghost pairs to   $(\beta^\alpha_r,\gamma^\alpha_r)^T$  
would give
$c=-48$.  Performing  compensating twists of the  fermionic fields   
in the four
packages appears to give matter fields with $c=48$.  However, there  
is some
redundancy -- the fields    $(\lambda_1^\alpha,\rho_1^\alpha)^T$  
can be grouped
with $(\beta_r,\gamma_r)^T$ in  a  topological package of  
$(1/2,1/2)$ fields
with vanishing central charge.  Likewise,  
$(\lambda_2^\alpha,\rho_2^\alpha)^T$
and the fields $(\delta^\alpha,\epsilon^\alpha)$  form a second  
topological
package of $(1/2,1/2)$ fields.   The residual matter has $c=40$  
that cancels
the ghosts.   Again the target space theory is  zero dimensional with the
right-moving fields  compactified in a manner consistent with modular
invariance and giving  a net number of 26  zero modes, consistent with an
object embedded in 26 dimensions.

\section{Beyond $N=4$}
 We have seen that by suitable field redefinitions the fields of critical
$(N,N')$ superstring theories with $N,N' \ge 1$  can be  
reinterpreted as ghost
fields of the theory with large $(4,4)$ supersymmetry.   These field
redefinitions amount to finite marginal deformations that change  
the conformal
weights of individual  fields in a manner that maintains the  
vanishing of the
total central charge. In this way the ghost and matter fields of  
the sectors
with $0\le N<3$ emerge from the $c=0$ $N=4$ ghost system.   There is no
suggestion that these  field definitions which embed theories with lower
supersymmetry in ones with higher supersymmetry are unique.  But,  
although it
is easy to imagine other ways of packaging the fields together, the ones
described so far  in this paper give a particularly intriguing   
interpretation
in which target-space coordinates emerge from the geometry (the  
ghost fields)
of the large $N=4$ super world-sheet.  Since these target spaces  
include those
of the  $(2,1)$ and $(2,2)$ theories we are here uniting theories  
whose target
spaces describe space-time with those which describe world-volumes  
embedded in
space-time (and, indeed, those with target spaces that describe the
world-volumes of theories whose target spaces are world-volumes,  
and so on
...).

 However,  we have seen that  it  may be  necessary to add some   
topological
packages    to the purely gravitational  sector of the $N=4$  
theory.    One
possibly unifying observation is that  there are supergravities  
with $N>4$
which automatically contain such extra fields \cite{baulieu1}.  In  these
theories the  ghosts  are  simply
identified as antisymmetric tensor representations of $O(N)$.    
Their action is
an obvious generalization of the gauge-fixed pure world-sheet $N=4$  
theory  --
a linear  ghost system obtained by fixing the gauge in which all  
the gauge
fields of the extended two-dimensional local supergravity vanish.    
The central
charge   vanishes for all $N\ge 3$.

  Deformations of these theories, analogous to those described in  
this paper,
lead to theories with lower $N$, but with extra topological  
packages appended.
For example,  the following array indicates how the  $N=5$ ghost  
field content
can be described as the sum  of the  $N=4$ theory together with an  
anomaly-free
set of  eight bosonic and eight fermionic fields,
\vskip 0.1cm
\be
\matrix{  {\rm weight} \cr
             1 & \bullet  \cr
            1/2 & \bullet &\bullet &\bullet &\bullet &\circ  \cr
          0  &\bullet &\bullet &\bullet &\bullet &\bullet &\bullet &\circ
&\circ &\circ &\circ \cr
            -1/2  &\bullet &\bullet &\bullet &\bullet &\circ &\circ  
&\circ
&\circ &\circ &\circ \cr
            -1 & \bullet &\circ &\circ &\circ &\circ \cr
             -3/2 & \circ \cr}\nn
\ee
\vskip 0.2cm
The $\circ $'s and $\bullet $'s  indicate the $N=5$ ghosts where  
the $\bullet
$'s are the  ghosts of the $N=4$ (labelled in the first column).  
The $\circ $'s
indicate the sixteen fermionic and bosonic ghosts that, together  
with their
antighosts, can be deformed by anomaly-free twists into two topological
packages of conformal weight $(1,0)$ fields.
Thus, the field content of the $N=5$ theory is large enough that  
the non-chiral
 supersymmetry-reducing  marginal deformations described earlier  
produce the
$(2,2)$ theory depicted in table 1, and hence the theories  with  
$N=2,1$ or $0$
supersymmetry   -- all the \lq matter' emerging from deformations  
of the ghost
fields.    We saw in section 3.3 that extra packages are needed to obtain
$(4,N)$ ($N\le 2$)  heterotic theories -- these extra fields could  
be obtained
by considering theories with  ever-increasing values of $N$.  The  
structure of
the $(4,N)$ theories is particularly intriguing and they surely  
merit further  investigation.

  In this paper we have discussed some hints that string  
world-sheets and their
target spaces
may be related in an underlying theory. These ideas come from the  
study of
deformations of theories based on two-dimensional world-sheets with  
extended
supersymmetry.
Since the \lq matter fields' of $N<3$  theories emerge in this  
procedure as
deformations of the
gravitational ghosts of theories with $N\ge 3$, we see that the  
target space
geometry may be encoded in some larger framework that includes  
world-sheet
structure.

\vskip 0.4cm
\noindent{\bf Acknowledgments}:\hfill\break\noindent
MBG is grateful  to the University of Paris VI and ER to the  
University of
Swansea  for periods of hospitality while this research was being  
carried out.
This work was supported in part by EEC under the TMR contract
ERBFMRX-CT96-0090.

\newcommand{\NP}[1]{Nucl.\ Phys.\ {\bf #1}}
\newcommand{\PL}[1]{Phys.\ Lett.\ {\bf #1}}
\newcommand{\CMP}[1]{Comm.\ Math.\ Phys.\ {\bf #1}}
\newcommand{\PR}[1]{Phys.\ Rev.\ {\bf #1}}
\newcommand{\PRL}[1]{Phys.\ Rev.\ Lett.\ {\bf #1}}
\newcommand{\PTP}[1]{Prog.\ Theor.\ Phys.\ {\bf #1}}
\newcommand{\PTPS}[1]{Prog.\ Theor.\ Phys.\ Suppl.\ {\bf #1}}
\newcommand{\MPL}[1]{Mod.\ Phys.\ Lett.\ {\bf #1}}
\newcommand{\IJMP}[1]{Int.\ Jour.\ Mod.\ Phys.\ {\bf #1}}
\newcommand{\JP}[1]{Jour.\ Phys.\ {\bf #1}}
\newcommand{\JMP}[1]{Jour.\ Math.\ Phys.\ {\bf #1}}
 
 \end{document}